\documentclass{article}

\usepackage{arxiv}
\usepackage[utf8]{inputenc} 
\usepackage[T1]{fontenc}    
\usepackage{hyperref}       
\usepackage{url}            
\usepackage{booktabs}       
\usepackage{amsfonts}       
\usepackage{nicefrac}           
\usepackage{lipsum}
\usepackage{graphicx}
\usepackage{colortbl}
\usepackage{xcolor}
\usepackage{array}
\usepackage{amsmath}
\usepackage{setspace}
\usepackage{appendix}
\graphicspath{ {./images/} }
\title{Instant prediction of relaxation in moir\'{e} superlattices using neural networks}

\author{
  Aleksei V. Belonovskii$^{1,*}$, Elizaveta I. Girshova$^1$, Erkki Lähderanta$^1$, Mikhail Kaliteevski \\
  \\
  $^1$Lappeenranta University of Technology LUT, Skinnarilankatu 34, 53850, Lappeenranta, Finland \\
  \\
  $^*$Corresponding author: \texttt{aleksei.belonovskii@gmail.com}
}

\begin{document}
\maketitle
\begin{abstract}
The relaxation of moir\'{e} superlattices in twisted bilayers of transition metal dichalcogenides (TMDs) has been modeled using a set of neural-network-based approaches. We implemented and compared several architectures, including (i) an interpolator combined with an autoencoder, (ii) an interpolator combined with a decoder, (iii) a direct generator mapping input parameters to displacement fields, and (iv) a physics-informed neural network (PINN). Among these, the direct generator architecture demonstrated the best performance, achieving machine-level precision with minimal training data. Remarkably, once trained, this simple fully connected network is able to predict the full displacement field of a moir\'{e} bilayer within a fraction of a second, whereas conventional continuum simulations require hours or even days. This finding highlights the low-dimensional nature of the relaxation process and establishes neural networks as a practical and efficient alternative to ab initio approaches for rapid modeling and high-throughput screening of 2D twisted heterostructures.
\end{abstract}


\section{Introduction}
Historically, the development of machine learning has been constrained by limited computational resources \cite{hutter2005universal, banko2001scaling, vapnik1998statistical} . Although a solid theoretical foundation for implementing algorithms had already been established in the last century, practical applications remained extremely limited. It was only with the advent of greater computational power and the accumulation of large datasets that the large-scale development of neural network models became feasible \cite{ref:1,ref:2}.

Nevertheless, the data problem remains highly relevant. Even large-scale systems like GPT have already exhausted a significant portion of publicly available information for training \cite{ref:3}. This issue becomes even more pressing in scientific and engineering domains, where obtaining new data often requires costly experiments, high-precision simulations, and substantial computational effort.

One could say that all the “low-hanging informational fruits” have already been picked—easy-to-access data has been utilized. In newer and more specialized areas, even small amounts of data come at a disproportionately high cost. As a result, it is precisely in physics, engineering, and other scientific disciplines that training neural networks proves especially challenging and expensive—not due to a lack of ideas, but because of the difficulty in assembling a high-quality training corpus \cite{ref:4,meng2025physics,schmidt2024improving, chang2022towards}. 

However, today, unlike the situation at the beginning of the 21st century, we have new methods for increasing the efficiency of neural networks. Transformers \cite{ashish2017attention}, Conformal Prediction (appeared around 2005, actively developed since 2018) \cite{vovk2022conformal, tibshirani2023conformal} , Bayesian models \cite{bayes1991essay,neal2012bayesian, linka2025discovering}  and other methods allow building reliable models even with a limited amount of training data. Therefore, one of the key areas of development at the moment is the development of neural networks that can work in conditions of insufficient data. There are already a number of approaches that allow modelling complex processes with a limited training sample \cite{ref:5, ref:6}. For example, in our previous work with transformers, we used one-hot-embedding, which allowed us to significantly reduce the amount of data required without losing the quality of predictions\cite{ref:7}.

In the 1960s, Shepard showed that the structure of stimuli affects the ease of categorization: categories with shared features (weakly correlated, independent) are easier to learn, categories with integral features (strongly correlated, interdependent) are more difficult, since they require the integration of all features and attract more attention \cite{ref:8}.
When the first simple neural networks with one hidden layer  were developed, it turned out that they did not show selective attention and learned equally on categories with separable and integral features when the stimuli were low-dimensional (e.g. rectangles). This is at odds with human behaviour. Deep networks, due to multiple hidden layers and the ability to abstract features at different levels, naturally show selective attention and learn in a more human-like manner, even on simple (low-dimensional) stimuli. However, recent studies show that when high-dimensional and complex stimuli are used (e.g. realistic faces), even one-hidden-layer networks begin to show human-like behaviour - that is, they learn faster on separable structures than on integral ones\cite{ref:9}.

The human mind evolutionary adapted to thrive at understanding complex systems by reducing them to a limited set of fundamental “images” (e.g., objects, concepts, patterns) and “properties,” linked by common sense rules \cite{ref:10,ref:11}. Usually, this mental representation involves no more than a dozen key elements \cite{ref:12,ref:13}. This intuitive approach is a kin to an analytic formula or a differential equation—a compact and elegant representation that captures the essence of a system’s behaviour.

In sharp contrast, modern numerical modelling describes systems using very big data arrays—thousands or millions of discrete values coupled through simple equations and boundary conditions. Although powerful, this method is computationally expensive, often consuming substantial time and resources to simulate complex systems.

A third paradigm—neural networks (NNs)—has recently achieved remarkable success. Neural networks generate predictions by identifying regularities and similarities between new inputs and a large training corpus of previously solved problems. Their core mechanism entails constructing a large matrix of weight coefficients (“weights”) that connects inputs to outputs through many layers. The dimensionality of this matrix is enormous (e.g., billions of parameters for models such as ChatGPT), making the initial “training” process extraordinarily compute-intensive. Once trained, however, an NN can produce predictions almost instantaneously.

In physics, many problems lack simple analytic solutions and must therefore rely on large-scale numerical simulation. Yet even when a system is represented by millions of data points, a human expert can often describe its essential state using a surprisingly small number of images and properties. This observation suggests that optimal solutions to such problems require an appropriate synthesis of human expertise, direct numerical modelling, and neural networks.

Consider, for example, modelling the structure of twisted van der Waals crystals—a promising class of semiconducting materials   \cite{Weston2020, rosenberger2020, woods2021, yoo2019}. A full numerical simulation may require solving a system of equations for more than 100,000 points, a process that can take days. Training a conventional neural network to predict the structure directly from atomic coordinates would be effective but impractical, because the weight matrix required would demand months of computation.

By contrast, the structure can be compressed into a human-interpretable form using key “images,” such as domain boundaries and their intersection points, and “properties,” such as divergence and curl within these domains  \cite{ Enaldiev_PRL, Mike_NanoLetters}. The number of such elements is far smaller than the original data points.

Accordingly, a more efficient strategy is to train a compact neural network that matches fundamental input parameters (e.g., material types, twist angle) directly to this concise set of descriptive images and properties. Such a smaller network would require far less computational power and training time, while remaining a powerful and rapid predictive tool for understanding the physical system.

\section{Physical Model of Twisted Bilayers}
\label{sec:physical_model_of_twisted Bilayers}

We consider bilayer systems composed of two-dimensional transition metal dichalcogenide (TMD) layers, such as $MoS_2$ and $WSe_2$ etc \cite{Mike_NanoLetters, Isaac_NanoLetters}. When one layer is rotated with respect to the other by a small twist angle, a moiré pattern emerges due to the interference of atomic lattices. This structural reconstruction gives rise to the formation of periodic domains.
\begin{figure}[h]
  \centering
  \includegraphics[width=1.0\linewidth]{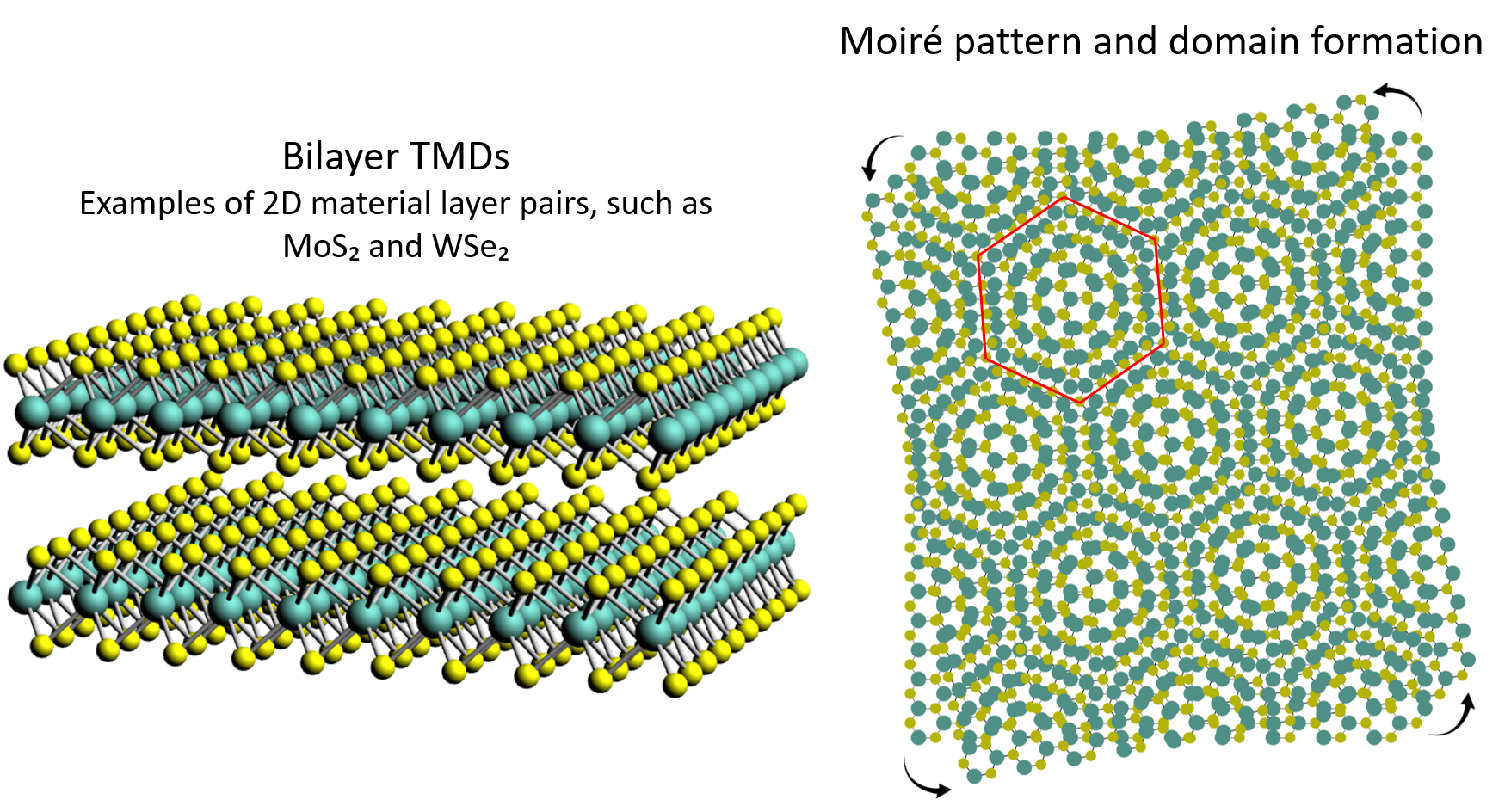}
  \caption{Formation of moiré superlattice and domains in twisted bilayers.}
  \label{fig:moire_domains}
\end{figure}

Figure~\ref{fig:moire_domains} illustrates a typical bilayer stacking (left) and the resulting moiré superlattice (right), which captures the essential geometric and physical features studied in this work.

When two atomic layers have similar lattice constants $a$ and $a'$ (where $\delta = (a' - a)/a \ll 1$), which can also be mutually twisted by a small angle $\theta \ll 1$, they produce a periodic moiré structure. The moiré lattice vector $\mathbf{l} $ relates to the crystalline lattice vector $\mathbf{a}$  through:

\begin{equation}
\mathbf{l} + \mathbf{a} = 
\begin{pmatrix}
1 + \delta & -\theta \\
\theta & 1 + \delta
\end{pmatrix} \mathbf{l}
\end{equation}

By rearranging terms, we obtain a more direct expression for the moiré vector:

\begin{equation}
\mathbf{a} = 
\begin{pmatrix}
\delta & -\theta \\
\theta & \delta
\end{pmatrix} \mathbf{l}
\end{equation}

This transformation matrix has complex eigenvalues, indicating a combination of scaling and rotation:

\begin{equation}
\lambda = \delta \pm i\theta
\end{equation}

The magnitude relationship and angular orientation follow directly from these eigenvalues:

\begin{align}
|\mathbf{a}| &= \sqrt{\delta^2 + \theta^2}  |\mathbf{l}| \\
\alpha &= \arctan\left(\theta / \delta\right)
\end{align}
where $\alpha$ represents the angle between the lattice vector $\mathbf{a}$ and the resulting moiré vector $\mathbf{l}$.

For identical layers with no lattice mismatch ($\delta = 0$), 
the eigenvalues become purely imaginary, confirming the rotational character of the transformation:

\begin{equation}
\lambda = \pm i\theta
\end{equation}

Consequently, the angle between vectors is exactly $90^\circ$:

\begin{equation}
\alpha = \pi/2
\end{equation}

The relaxed moiré superlattice structure in twisted TMD bilayers results from competition between elastic and adhesion forces. Adhesion favors formation of natural stacking parallel (P) or antiparallel (AP)  domains. Elastic forces resist deformation from the bare monolayer lattice constants.

The total energy functional combines adhesion and elastic contributions:

\begin{equation}
E = \int dr^2 \, \left[ e_{\text{elastic}} + W_{\mathrm{AP}/\mathrm{P}} \right]
\end{equation}
where \\
\begin{tabular}{p{5cm}l}
$e_{\text{elastic}}$& elastic energy density \\
$W_{\mathrm{AP}/\mathrm{P}}$ & adhesion energy density\\
\end{tabular}

\medskip

The elastic energy density per layer decomposes as:
\begin{equation}
e_{\text{elastic}}^{(l)} = 
\underbrace{
  \frac{\lambda_l + \mu_l}{2} \left( \mathrm{div}\ \mathbf{u}^{(l)} \right)^2
}_{\text{Hydrostatic strain energy}} 
+ 
\underbrace{
  \frac{\mu_l}{2} \left[ 
    \left( u_{xx}^{(l)} - u_{yy}^{(l)} \right)^2 + 
    4 \left( u_{xy}^{(l)} \right)^2 
  \right]
}_{\text{Shear strain energy}}
\end{equation}

where \\
\begin{tabular}{p{5cm}l}
$\lambda_l, \mu_l$ & elastic moduli for layer $l$ (W or Mo) \\
$\mathbf{u}^{(l)}$ & displacement field in layer \\
$ u_{ij}^{(l)} = \frac{1}{2}(\partial_i u_j^{(l)} + \partial_j u_i^{(l)})$ & strain tensor components \\
$\mathrm{div}\  \mathbf{u}^{(l)} = \partial_x u_x^{(l)} + \partial_y u_y^{(l)}$ & hydrostatic strain
\end{tabular}

\medskip
For adhesion energy calculation we need to define an in-plane vector $\vec{r_0}$ determining the stacking arrangement between layers:
\begin{align}
\mathbf{r}_0(\mathbf{r}) = \delta \cdot \mathbf{r} + \theta \, \hat{z} \times \mathbf{r} 
+ \mathbf{u}^{\text{t}}(\mathbf{r}) - \mathbf{u}^{\text{b}}(\mathbf{r}),
\end{align}

where

\begin{tabular}{p{5cm}l}
 $\delta$  & lattice mismatch parameter ($\delta \approx 0.4\%$ for MoSe$_2$/WSe$_2$)\\
$\theta$ & twist angle between the layers \\
$\mathbf{u}^{\text{t}}(\mathbf{r})$ and $\mathbf{u}^{\text{b}}(\mathbf{r})$ & in-plane displacement fields \\
\end{tabular}

\medskip

The adhesion energy is given by:
\begin{align}
W_{P/AP}(r_0) = & -\varepsilon Z^2(r_0) 
+ w_1 \sum_{n=1,2,3} \cos \big( G^{(1)}_n r_0 \big) \nonumber \\
& + w_2 \sum_{n=1,2,3} \sin \big( G^{(1)}_n r_0 + \gamma_{P/AP} \big).
\end{align}

\begin{align}
Z(r_0) = \frac{1}{2\varepsilon} \sum_{n=1}^{3} \Big[ 
  & w_1 \sqrt{G^2 + \rho^{-2}} \cos \big( G^{(1)}_n r_0 \big) \nonumber \\
  & + w_2 G \sin \big( G^{(1)}_n r_0 + \gamma_{P/AP} \big) \Big],
\end{align}

where \\
\begin{tabular}{p{2cm}l}
$w_n$ & interaction amplitudes, $w_1 = A_1 e^{-d_0 \sqrt{G^2 + \rho^{-2}}}$, 
$w_2 = A_2 e^{-d_0 G}$\\
$A_n$ & adhesion coefficients (See Table 1)\\
$d_0$& interlayer distance (See Table 1)\\
$\varepsilon$& effective stiffness (See Table 1)\\
$\mathbf{G}_n^{(k)}$ & reciprocal lattice vectors for harmonic $k$ and direction $n=1,2,3$\\
$\rho$ & decay length of the adhesion potential in reciprocal space (See Table 1)\\
$\gamma_{\mathrm{P}}$ & $\pi/2$ for parallel orientation\\
$\gamma_{\mathrm{AP}}$ & $  0$ for antiparallel orientation
\end{tabular}

\medskip

\bigskip

\begin{table}[h!]
\centering
\caption{Fitting parameters for adhesion energy.}
\begin{tabular}{lccccc}
 & $A_1$, eV/nm$^2$ & $A_2$, eV/nm$^2$ & $\rho$, nm & $d_0$, nm & $\varepsilon$, eV/nm$^4$ \\
MoS$_2$/MoS$_2$ & 71928800 & 56411 & 0.0496 & 0.65 & 214 \\
MoTe$_2$/MoTe$_2$ P & 1660909 & 53254 & 0.0162 & 0.742 & 219 \\
MoTe$_2$/MoTe$_2$ AP & 1327437 & 108134 & 0.0162 & 0.742 & 219 \\
WS$_2$/WS$_2$ & 84571600 & 70214 & 0.0495 & 0.65 & 213 \\
WSe$_2$/WSe$_2$ & 121287200 & 110873 & 0.0497 & 0.69 & 190 \\
MoSe$_2$/MoSe$_2$ & 96047400 & 81488 & 0.0506 & 0.68 & 189 \\
\end{tabular}
\end{table}

From these definitions, we can obtain the expressions for the parallel and antiparallel configurations:

\begin{multline}
W_P(\mathbf{r}_0) = \left[ w_1 + w_2 - \frac{\left( w_1 \sqrt{G^2 + \rho^{-2}} + w_2 G \right)^2}{4\varepsilon} \right] \sum_{n=1}^{3} \cos \left( \mathbf{G}_n^{(1)} \cdot \mathbf{r}_0 \right) \\
- \frac{\left( w_1 \sqrt{G^2 + \rho^{-2}} + w_2 G \right)^2}{4\varepsilon} \left[ \sum_{n=1}^{3} \cos \left( \mathbf{G}_n^{(2)} \cdot \mathbf{r}_0 \right) + \frac{1}{2} \sum_{n=1}^{3} \cos \left( \mathbf{G}_n^{(3)} \cdot \mathbf{r}_0 \right) \right]
\end{multline}

\begin{multline}
W_{AP}(\mathbf{r}_0) = \left[ w_1 - \frac{\left( w_1 \sqrt{G^2 + \rho^{-2}} \right)^2 - (w_2 G)^2}{4\varepsilon} \right] \sum_{n=1}^{3} \cos \left( \mathbf{G}_n^{(1)} \cdot \mathbf{r}_0 \right) \\
- \frac{\left( w_1 \sqrt{G^2 + \rho^{-2}} \right)^2 + (w_2 G)^2}{4\varepsilon} \sum_{n=1}^{3} \cos \left( \mathbf{G}_n^{(2)} \cdot \mathbf{r}_0 \right) \\
- \frac{\left( w_1 \sqrt{G^2 + \rho^{-2}} \right)^2 - (w_2 G)^2}{8\varepsilon} \sum_{n=1}^{3} \cos \left( \mathbf{G}_n^{(3)} \cdot \mathbf{r}_0 \right) \\
+ \left[ w_2 + \frac{w_1 w_2 G \sqrt{G^2 + \rho^{-2}}}{2\varepsilon} \right] \sum_{n=1}^{3} \sin \left( \mathbf{G}_n^{(1)} \cdot \mathbf{r}_0 \right) \\
- \frac{w_1 w_2 G \sqrt{G^2 + \rho^{-2}}}{4\varepsilon} \sum_{n=1}^{3} \sin \left( \mathbf{G}_n^{(3)} \cdot \mathbf{r}_0 \right)
\end{multline}

As a result, TMG twisted bilayer is transformed to Moire superlattice made of domain mimicking bulk crystlal (slightly strained), separated by boundaries, where strong strain are located, see \cite{Mike_NanoLetters} for the details. The parameters, defining adhesion energy for various bi-layers are summarized in Table 1.

\section{Neural Network Modeling Framework}
\label{sec:neural_network_modeling_framework}
To accelerate and generalize the prediction of displacement fields in twisted bilayer systems, we propose a family of neural network architectures capable of mapping structural parameters to full-field displacement solutions.

Each simulation produces a high-dimensional output (displacement matrices). The input consists of physical parameters such as the twist angle, stacking type, and material identities and their associated physical constants.

The goal of the neural network is to learn a surrogate model that efficiently predicts the displacement field given only a small set of physical descriptors, reducing computational cost and enabling generalization across materials and configurations.

The figure~\ref{fig:model} illustrates the pipeline of our machine learning framework. Given the twist angle, stacking type, and material descriptors, the model predicts the displacement field through a neural network. The output can be postprocessed to visualize displacement field.

It should be emphasized that the proposed framework is equally applicable to both homostructures and heterostructures, where twists may appear. Since this distinction does not affect the training procedure of the neural networks, in what follows we restrict our analysis to homostructures; the results for heterostructures are expected to follow the same trends.

We analyze the displacement fields arising in layered 2D materials with different stacking configurations and twist angles. The following material pairs are considered:

\begin{figure}[h]
  \centering
  \includegraphics[width=1.0 \linewidth]{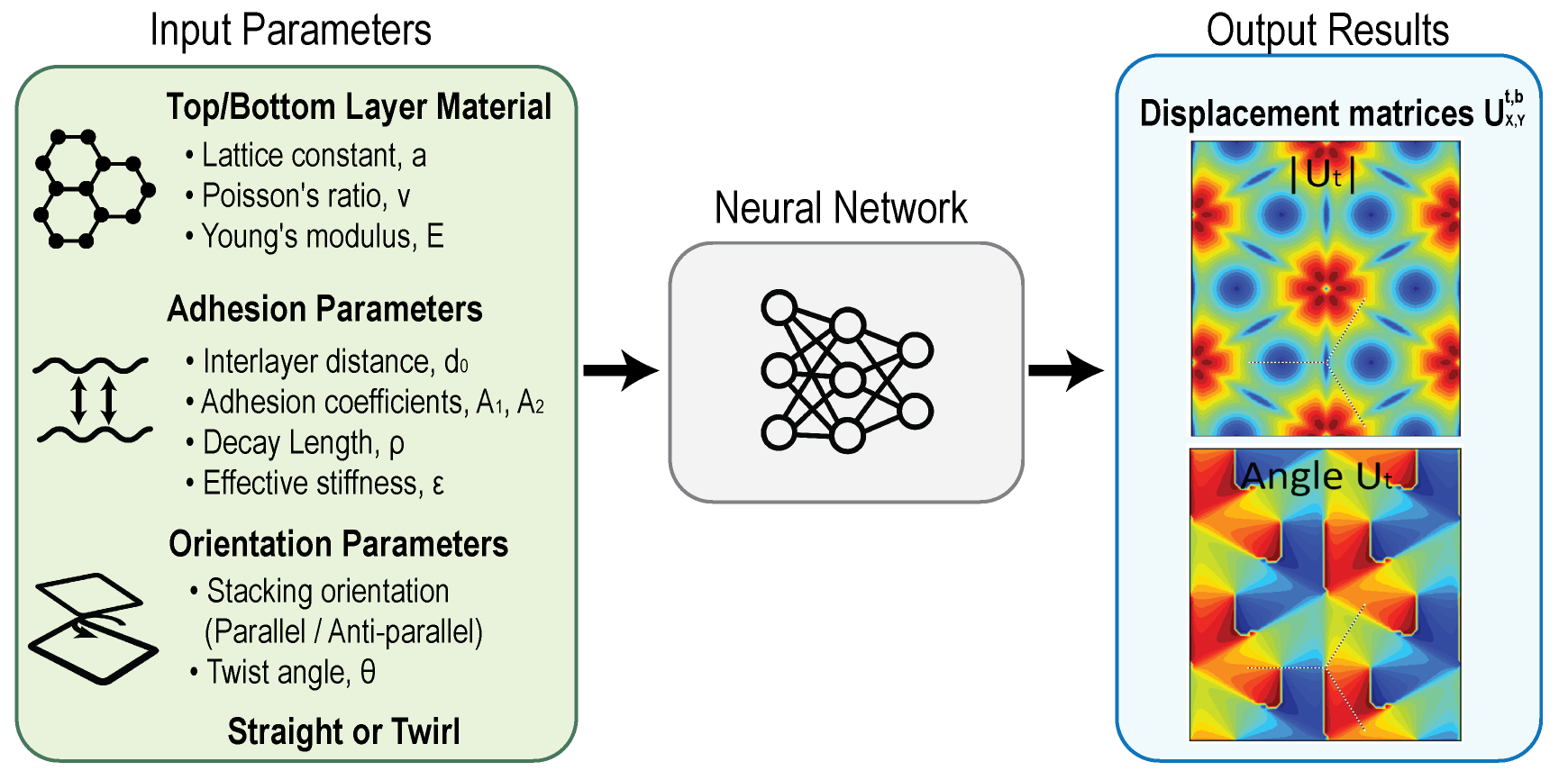}
  \caption{ Neural network modeling pipeline for twisted bilayers. Input parameters include twist angle, configuration, material constants and type of Chirality. The neural network maps these to a high-dimensional displacement field.}
  \label{fig:model}
\end{figure}

We analyze the displacement fields arising in layered 2D materials with different stacking configurations and twist angles. The following material pairs are considered:

\begin{table}[h!]
\centering
\begin{tabular}{|l|c|c|}
\hline
\textbf{Material} & \textbf{Stackings} \\
\hline
\textbf{MoTe\textsubscript{2}–MoTe\textsubscript{2}} & P, AP \\
\textbf{WS\textsubscript{2}–WS\textsubscript{2}} & P, AP  \\
\textbf{MoS\textsubscript{2}–MoS\textsubscript{2}} & P, AP \\
\textbf{WSe\textsubscript{2}–WSe\textsubscript{2}} & P, AP  \\
\textbf{MoSe\textsubscript{2}–MoSe\textsubscript{2}} & P, AP \\
\hline
\end{tabular}
\caption{Considered homostructures with parallel (P) and antiparallel (AP) stackings.}
\end{table}

For each configuration, simulations are performed for \textbf{200 angles}, ranging from \textbf{0.01\textdegree{} to 2.00\textdegree{}} with a step of \textbf{0.01\textdegree{}}.

These generate 4 displacement matrices per simulation: \texttt{utx}, \texttt{uty}, \texttt{ubx}, \texttt{uby}, which are later flattened and concatenated to form a single high-dimensional vector.

Additionally, visualizations of these results (e.g., angle dependence) are available in the form of combined moiré videos.

Fixed parameters: $N_x = 39$ and $N_y = 39$.

\begin{figure}[h]
  \centering
  \includegraphics[width=1.0\linewidth]{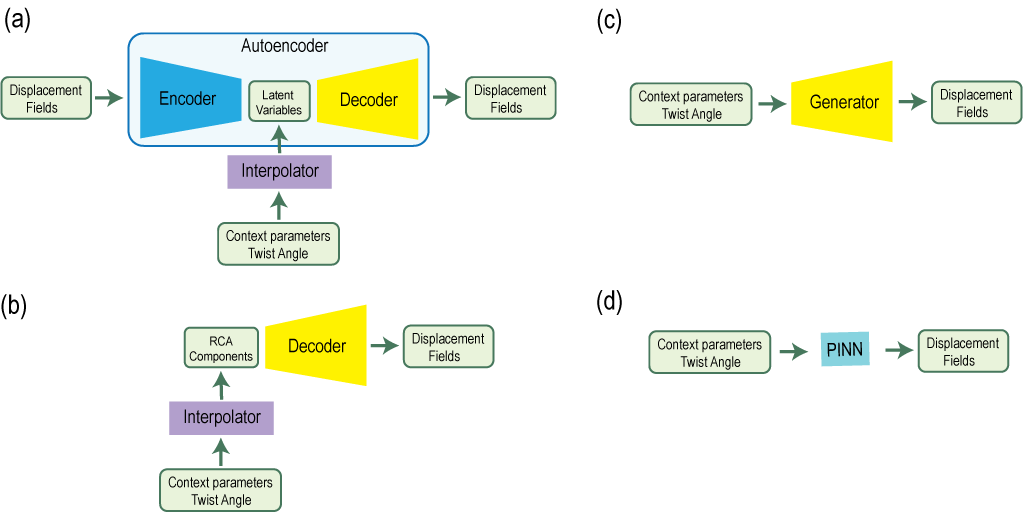}
  \caption{Four neural network architectures for predicting displacement fields:  
  (a) Interpolator + Autoencoder;  
  (b) Interpolator + Decoder;  
  (c) Direct generator from context to displacement fields;  
  (d) Physics-Informed Neural Network solving the governing equations directly.}
  \label{fig:nn_architectures}
\end{figure}

Figure~\ref{fig:nn_architectures} presents four neural network architectures designed to model displacement field distributions in bilayer 2D material systems.

\begin{itemize}
  \item[(a)] \textbf{Interpolator + Autoencoder.}  
  The first scheme shows a model based on an autoencoder. The encoder compresses the input displacement fields into a compact latent space consisting of a few physically meaningful variables. Simultaneously, an interpolator is trained to predict the same latent variables based on input parameters — namely, the material context and twist angle.

  In the final configuration, the interpolator takes the context and twist angle as input, predicts the latent vector, and the decoder reconstructs the full displacement field.  
  This approach shifts the computational burden of generation to the decoder and reduces data requirements.  
  The encoder, decoder, and interpolator are trained jointly using a composite loss function, encouraging the encoder to extract latent representations that are both physically interpretable and useful for interpolation and reconstruction.

  \item[(b)] \textbf{Interpolator + Decoder.}  
  In the second architecture, the encoder is omitted. Instead, latent variables are replaced with RCA (Reduced Component Analysis) parameters obtained via prior analysis. These are directly fed into the decoder.  
  This approach reduces training time, as no encoder is needed, and RCA components are known to effectively describe system behavior. However, it relies on precomputed RCA results. The interpolator is trained to predict RCA parameters from context.

  \item[(c)] \textbf{Direct Generator.}  
  The third model is a simple fully connected neural network that directly maps input parameters (context and angle) to the displacement fields.  
  This architecture offers maximum flexibility, as it learns the entire mapping without any predefined latent structure or assumptions.

  \item[(d)] \textbf{Physics-Informed Neural Network (PINN).}  
  The fourth architecture utilizes a Physics-Informed Neural Network (PINN), adapted for solving nonlinear systems of equations that govern the displacement behavior.  
  The network receives physical parameters that define the equations and learns to solve them by minimizing the residual of the system rather than comparing with labeled output.  
  Unlike the previous approaches, PINNs do not require precomputed datasets and are ideal for cases where simulations are expensive but a physical model is known.
\end{itemize}

In Figure~\ref{fig:nn_architectures}, we separated the input parameters into \textit{angle} and \textit{context parameters}, where the angle can vary quite freely with many possible configurations. The context parameters, such as material properties, exhibit less variation since they depend on the specific material types considered in this study, which are limited in number.

In the case of Physics-Informed Neural Networks (PINNs), the learning objective fundamentally differs from standard supervised training: the true solution \( x_{\text{true}} \) is unknown. Instead, the network is trained to produce a predicted solution \( x_{\text{pred}} \), which is then substituted into the governing equation \( F \) to evaluate how strongly the physical model is violated. This residual defines the learning signal. The general form of the equation can be written as:
\[
F(x_{\text{pred}}) = A x_{\text{pred}} + f(x_{\text{pred}}) \neq 0,
\]
and the loss function used to guide optimization is based on minimizing the residual norm:
\[
\text{Loss} = \| F(x_{\text{pred}}) \|^2.
\]

Backpropagation in PINNs requires a nested chain of derivatives, since the loss depends on the residual, which in turn depends on the predicted solution, which itself depends on the network parameters. This is expressed as:
\[
\frac{d\, \text{Loss}}{d \theta} = \frac{d\, \text{Loss}}{d F} \cdot \frac{d F}{d x} \cdot \frac{d x}{d \theta},
\]
where \( \theta \) denotes the parameters of the neural network. In our specific case, the derivative simplifies to:
\[
\frac{d\, \text{Loss}}{d \theta} = 2 F \cdot J \cdot \frac{d x}{d \theta},
\]
where \( J = \frac{d F}{d x} \) is the Jacobian matrix of the system.

This process is substantially more computationally expensive than standard supervised learning, for several reasons:

\begin{itemize}
  \item The ground truth solution \( x_{\text{true}} \) is unavailable;
  \item Each training step requires evaluating the residual by substituting the predicted vector \( x_{\text{pred}} \) into the nonlinear system \( F \);
  \item Backpropagation must pass through matrix operations involving the Jacobian \( J \).
\end{itemize}

As a result, the training process converged slowly, and the loss rarely fell below 0.01.

To improve the learning signal, we explored an alternative formulation based on \textit{Newton's method}. Assuming that the scale of the residual \( F \) and the error \( x_{\text{true}} - x_{\text{pred}} \) might differ significantly, we evaluated the norm of the Newton update step as a surrogate for the learning objective:
\[
x_{i+1} = x_i - J^{-1} F.
\]
Accordingly, the loss was redefined as the squared norm of the Newton step:
\[
\text{Loss} = \| \delta x \|^2 = \| -J^{-1} F \|^2,
\]
under the assumption that the Newton step \( \delta x \) may correlate more directly with the actual prediction error.

In this formulation, the gradient of the loss with respect to the network parameters becomes significantly more complex:
\[
\frac{d\, \text{Loss}}{d \theta} = 2 \left( J^{-1} F \right) \cdot \left( J^{-1} \frac{d J}{d x} J^{-1} F - 1 \right) \cdot \frac{d x}{d \theta}.
\]

While theoretically promising, this approach proved computationally expensive: a single epoch under this Newton-step-based formulation could take up to 30 minutes, rendering the training process largely impractical for large datasets or real-time inference.

\section{Results and Discussion}
\label{sec:results_discussion}
Before comparing the performance of these neural architectures, we perform an RCA-based analysis of the displacement fields.  
This analysis helps determine the number of latent variables required to adequately describe the system and is directly used in architecture (b), while also serving as a reference for the learned representations in architecture (a).

To analyze the intrinsic dimensionality of the displacement data, we apply Principal Component Analysis (PCA) to the flattened displacement vectors. Let $\mathbf{X}$ denote the data matrix of shape $(200, D)$, where each row corresponds to the concatenated displacement fields for a specific twist angle, and $D$ is the total number of flattened features (e.g., $D = 4 \times 39 \times 39 = 6084$).

As an illustrative example, the PCA results for the \texttt{MoTe\textsubscript{2}–MoTe\textsubscript{2}} homostructure in parallel (P) stacking configuration are shown below:

\begin{itemize}
  \item Component 1: 92.90\%
  \item Component 2: 6.09\%
  \item Component 3: 0.68\%
  \item Component 4: 0.23\%
  \item Component 5: 0.07\%
  \item Component 6: 0.02\%
  \item Component 7: 0.01\%
  \item Components 8--10: $\leq 0.01$\%
\end{itemize}

The first two principal components capture over 98\% of the variance in the data. This indicates that the displacement fields, despite being high-dimensional, effectively lie on a two-dimensional manifold. This insight significantly simplifies subsequent modeling and interpolation tasks.

\vspace{1em}

\begin{figure}[h]
  \centering
  \includegraphics[width=1.0\linewidth]{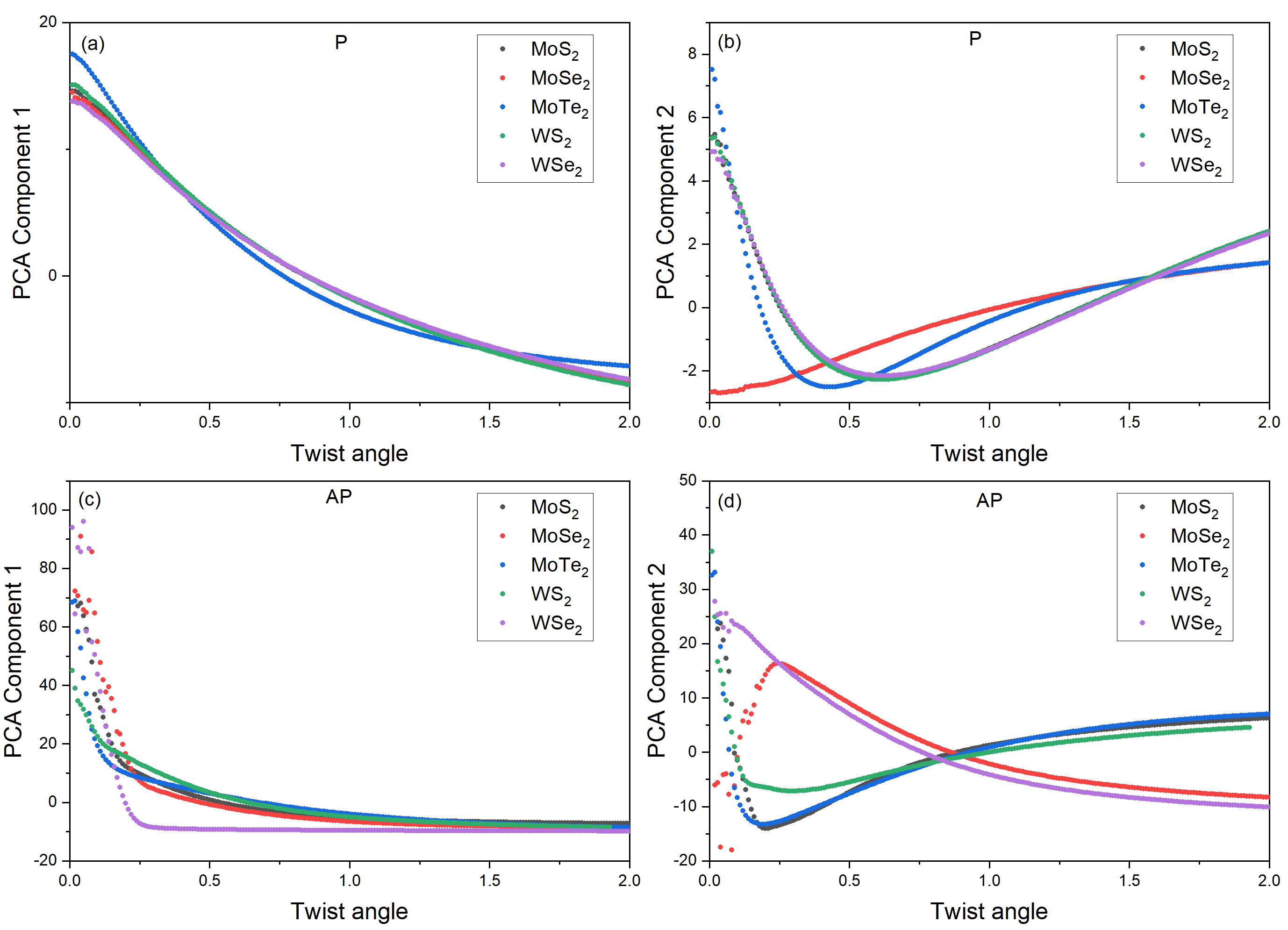}
  \caption{Principal Component 1 and 2 as functions of twist angle $\theta$ for five homostructures:  
  (a) Component 1 for P orientation;  
  (b) Component 2 for P orientation;  
  (c) Component 1 for AP orientation;  
  (d) Component 2 for AP orientation.  
  PCA components exhibit smooth angle dependence, supporting their use in interpolation schemes.}
  \label{fig:pca_angle_dependence}
\end{figure}

Figure~\ref{fig:pca_angle_dependence} shows the dependence of the first two PCA components on the twist angle $\theta$ for all considered homostructures. Subfigures (a) and (b) correspond to the parallel (P) stacking configuration, while (c) and (d) show the antiparallel (AP) configuration. 

It can be observed that the first principal component, which explains more than 90\% of the total variance, exhibits a relatively smooth and stable dependence across different structure pairs. This suggests that its variation is dominated by a single key factor—namely, the incidence angle.

Based on this observation, we will use the angle as the primary input parameter during training. The remaining parameters, which are influenced by material properties and structural orientation, will be treated as contextual information. These context-dependent features will be incorporated through lightweight adaptation for each material type, using a strategy based on task-specific \textit{fine-tuning}. This setup allows the model to generalize across materials while still capturing material-specific nuances where needed.

\begin{figure}[htbp]
    \centering
    \includegraphics[width=1.0\linewidth]{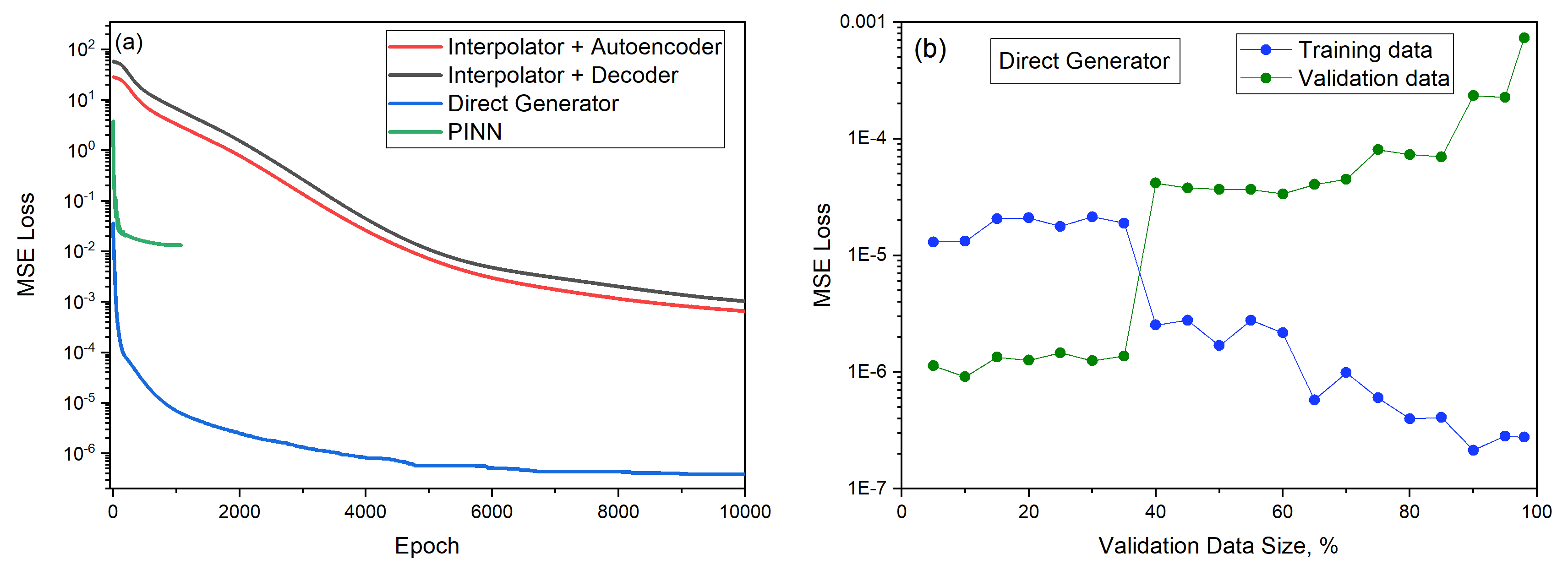}
    \caption{Model performance comparison.  
    (a) Validation MSE vs. epochs for all considered models using 80\% training and 20\% validation data.  
    (b) Validation and training error of the Direct Generator as a function of training set size.  
    The Direct Generator shows the best performance and generalization even under extreme data scarcity, while PINN exhibits slower convergence and significantly higher error.}
    \label{fig:figure5}
\end{figure}

Figure ~\ref{fig:figure5}a illustrates the training performance of all considered architectures. The plots show the validation mean squared error (MSE) as a function of training epochs, using a standard split of 80\% for training and 20\% for validation.  
The Interpolator + Autoencoder and Interpolator + Decoder architectures exhibit similar convergence behavior. The former achieves a final validation error of $6 \times 10^{-4}$, while the latter converges to approximately $1 \times 10^{-3}$ by epoch 10{,}000. The slight advantage of the autoencoder-based approach may stem from its flexible latent representation, which is not restricted by a fixed RCA basis.

The best performance is observed with the Direct Generator model. It converges significantly faster, reaching an MSE of $1 \times 10^{-6}$ by epoch 3000, and ultimately achieving a validation error of $3.86 \times 10^{-7}$ at epoch 10{,}000 — approaching machine precision.

In contrast, the Physics-Informed Neural Network (PINN) performs notably worse, with a validation error plateauing at 0.013. Moreover, PINN training is computationally intensive. While training a standard network takes milliseconds per epoch, PINN training — especially with Newton steps — can require up to 30 minutes per epoch due to matrix operations required to evaluate residuals of the nonlinear system. As a result, PINN training was halted after 1000 epochs when no further improvements were observed.  
No significant difference was found between using Newton-step losses and standard residual loss.

Figure ~\ref{fig:figure5}b explores model generalization across varying training set sizes for the Direct Generator. The following regimes can be observed:
\begin{itemize}
  \item Up to 40\% validation data, the validation loss remains consistently lower than the training loss.
  \item Between 40--50\%, the validation and training curves intersect, marking a transition zone.
  \item From 50\% to 85\%, both losses remain low and close, indicating stable generalization.
  \item Beyond 85\%, the validation error increases rapidly due to overfitting, as the model is trained on very few examples.
\end{itemize}

Nevertheless, even with only 4 training examples (98\% validation), the generator still achieves a validation error of $7 \times 10^{-4}$. However, at 99\%, the error increases significantly, indicating the limits of generalization under extreme data scarcity.

\begin{figure}[htbp]
    \centering
    \includegraphics[width=1.0\linewidth]{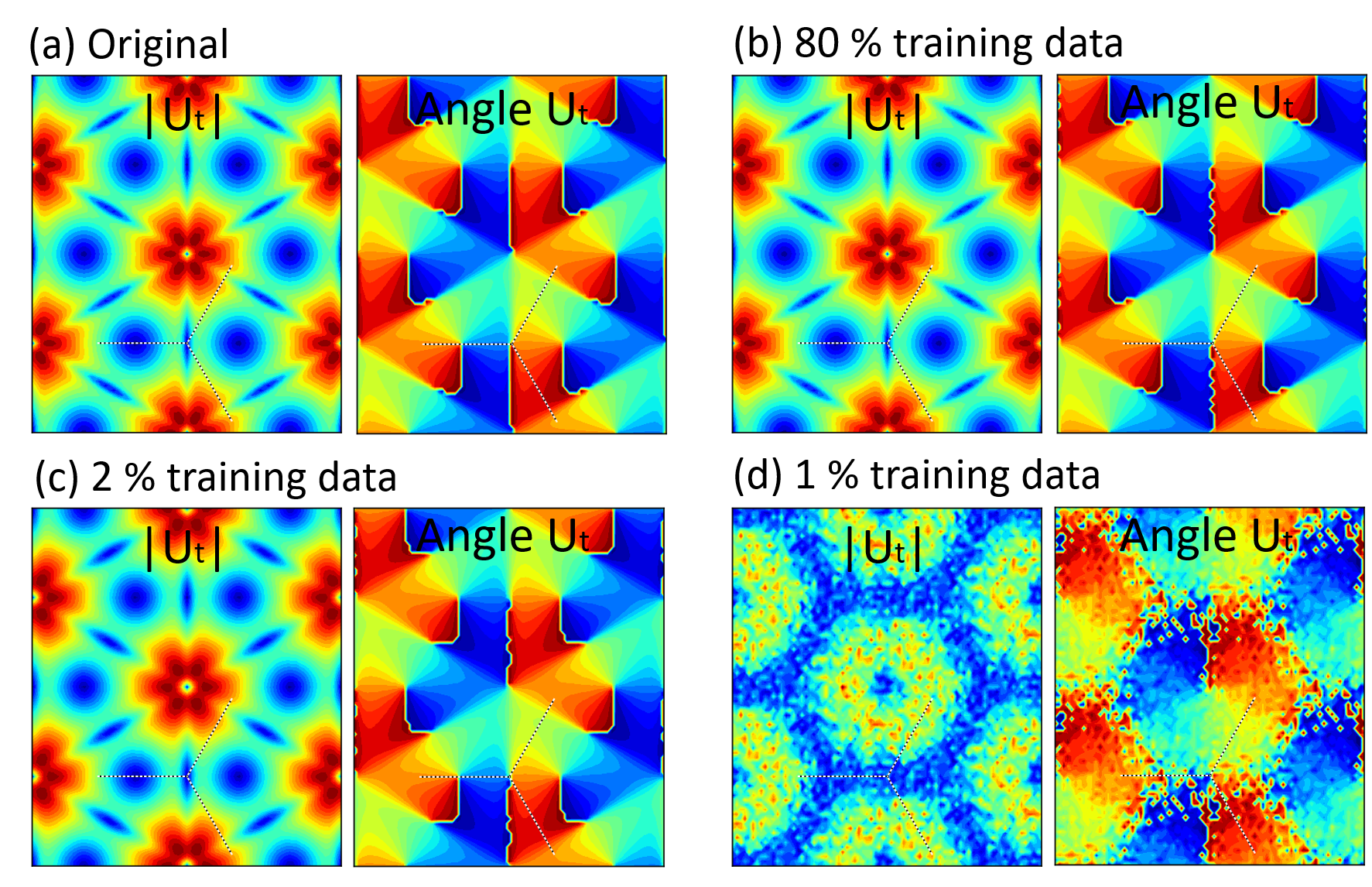}
    \caption{Displacement field predictions generated by the Direct Generator under different training data conditions.  
    (a) Ground truth (reference simulation).  
    (b) Prediction using 160 training samples (80\% of data for training).  
    (c) Prediction using 4 training samples (2\% of data for training).  
    (d) Prediction using 2 training samples (1\% of data for training).}
    \label{fig:figure6}
\end{figure}

Figure~\ref{fig:figure6} shows qualitative reconstruction results for displacement fields predicted by the Direct Generator under different training set sizes.  
These results are obtained for a fixed input configuration: a MoTe–MoTe bilayer system with parallel (P) orientation and a twist angle of $0.1^\circ$.  
Only the Direct Generator model is evaluated in this comparison, as it demonstrated the highest generalization performance in previous tests.

\begin{itemize}
  \item[(a)] Ground truth (reference simulation).
  \item[(b)] Predicted with 80\% training data.
  \item[(c)] Predicted with 2\% training data (only 4 training samples).
  \item[(d)] Predicted with 1\% training data (only 2 training samples).
\end{itemize}

The reconstructed fields remain visually close to the reference even under severe data constraints, highlighting the strong generalization capabilities of the Direct Generator model. Despite the extreme data scarcity in cases (c) and (d), the model is able to produce fields that remain visually and quantitatively close to the ground truth.  
This result demonstrates that even a simple fully connected network can be successfully trained using as few as four examples.

The conducted study demonstrates that neural networks, especially the Direct Generator architecture, offer a powerful and efficient alternative to traditional ab initio methods for modelling relaxation and predicting displacement fields in twisted 2D homostructures.

The key result is the ability of even a simple fully connected network (Direct Generator) to accurately reproduce complex displacement fields, learning from an extremely small amount of data - just a few examples. This indicates that the physics of the relaxation process, despite the high dimensionality of the original data, is effectively described by a low-dimensional manifold, which allows the neural network to generalize the process.

In contrast, PINN, although they do not require pre-computations for training, showed significantly slower convergence and worse accuracy in this problem, which is due to the computational complexity of handling non-linear systems of equations and calculating Jacobian matrices.

The practical meaning of this findings is the creation of a tool for instant prediction of the structure of a relaxed moiré grating (in seconds) after training, while traditional numerical methods require hours or days of calculations. This opens up opportunities for high-performance screening of materials and rapid study of the influence of the twist angle and other parameters on the properties of van der Waals homo- and heterostructures. In the future, the development of this approach may include the integration of more complex architectures (for example, transformers) to account for a wider range of materials and layer configurations, as well as combining the prediction speed of neural networks with the physical rigor of PINN in hybrid models.

\section{Conclusions}
\label{sec:conclusions}
In this work, we developed and compared several neural-network-based approaches for modeling relaxation in twisted bilayers of transition metal dichalcogenides. We demonstrated that the Direct Generator architecture, a simple fully connected network, achieves the highest accuracy and generalization capability.

The key result is that the Direct Generator reproduces full displacement fields with machine-level precision, even when trained on as few as four examples. This confirms that the relaxation process, though high-dimensional in its raw form, is governed by a low-dimensional manifold that neural networks can effectively capture.

In contrast, physics-informed neural networks (PINNs) converged slowly and showed reduced accuracy, reflecting the computational burden of handling nonlinear systems with Jacobians during training. While PINNs remain attractive for problems without reference data, their application to this task proved inefficient.

The practical implication of our findings is clear: once trained, a neural network can predict the relaxed moiré superlattice structure within seconds, in stark contrast to hours or days required by ab initio methods. This enables rapid high-throughput screening of 2D bilayer systems across a wide range of twist angles and material combinations.
 
\bibliographystyle{unsrt}  

\end{document}